\documentclass[aps,twocolumn,prx,longbibliography,floatfix,superscriptaddress,footnoteinbib,10pt]{revtex4-1}

\usepackage[dvips]{graphicx}
\usepackage{latexsym}
\usepackage{amsmath}
\usepackage{amssymb}
\usepackage{amsfonts}
\usepackage{color}
\usepackage{bm}
\usepackage{verbatim}
\usepackage[english]{babel}
\usepackage[utf8]{inputenc}
\usepackage{euscript}
\usepackage{subfigure}
\usepackage[normalem]{ulem}

\usepackage[colorlinks,bookmarks=false,citecolor=blue,linkcolor=blue,urlcolor=blue]{hyperref}
\usepackage[all]{hypcap} 

\graphicspath{{.},{../../pics/},{../../pics/draw/}}

\newcommand{\sign}{\mathop{\mathrm{sign}}}
\newcommand{\Tr}{\mathop{\mathrm{Tr}}}

\usepackage{ulem} 

\begin{document}
\title{
Higgs modes in proximized superconducting systems}

\author{V. L. Vadimov}
\affiliation{Institute for Physics of Microstructures, Russian
Academy of Sciences, 603950 Nizhny Novgorod, GSP-105, Russia}
\author{I. M. Khaymovich}
\affiliation{Max Planck Institute for the Physics of Complex Systems,01187 Dresden, N\"othnitzer Stra{\ss}e 38, Germany }
\author{A. S. Mel'nikov}
\affiliation{Institute for Physics of Microstructures, Russian
Academy of Sciences, 603950 Nizhny Novgorod, GSP-105, Russia}

\begin{abstract}

The proximity effect in hybrid superconducting - normal metal structures is shown to affect strongly the
 coherent oscillations of the superconducting {order parameter } $\Delta$ known as the Higgs modes.
The standard Higgs mode at the frequency $2\Delta$ is damped exponentially by the quasiparticle leakage from the primary superconductor.
{Two new Higgs modes with the frequencies depending on both the primary and induced gaps in the hybrid structure are shown to appear due to
t}he coherent electron transfer between the superconductor and the normal metal. Altogether these three modes determine the long--time asymptotic behavior of the superconducting order parameter disturbed either by the
electromagnetic pulse or the quench of the system parameters and, thus, are of crucial importance for the dynamical properties and restrictions on the operating frequencies for superconducting devices based on the proximity effect used, e.g., in quantum computing, in particular, with topological low-energy excitations.
\end{abstract}

\maketitle


The progress of modern nanotechnology opens new horizons
for engineering superconducting correlations in various hybrid structures
and creating, in fact, novel types of artificial superconducting materials
with controllable
properties~\cite{oreg2010helical,qi2011topological,sato2017topological,
    heersche2007induced,sato2008gate,kopnin2013predicted,kopnin2013vortex,lee2018proximity,
    lutchyn2018majorana, aasen2016milestones, alicea2012new}.
The proximity phenomenon arising in a non-superconducting material from the electron exchange with a primary superconductor
can generate the induced superconducting ordering in a wide class of materials, including unconventional ones~\cite{oreg2010helical,qi2011topological,sato2017topological,
    heersche2007induced,sato2008gate,lee2018proximity,
    lutchyn2018majorana, aasen2016milestones, alicea2012new}.
The resulting superconducting state in these materials can controllably reveal the exotic properties
very rarely found in natural metals or alloys and strongly different from the ones of the primary superconductor.
The induced Cooper pairs can
change, e.g., their spin structure from the singlet to a triplet one in the
presence of strong spin-orbit coupling and Zeeman (or exchange)
field~\cite{oreg2010helical,lutchyn2018majorana, aasen2016milestones}
This spin transformation affects,
of course,  the momentum space structure of pairs:
the routine s-wave condensate can turn into an exotic p-wave one.
The resulting Cooper pair structure leads to the formation of topological
low-energy excitations such as Majorana fermions~\cite{oreg2010helical,
lutchyn2018majorana, aasen2016milestones, alicea2012new}
and possesses a high potential for
the development of new types of nanoelectronic devices perspective for
applications in quantum computing, quantum information processing, quantum
annealing, quantum memory and others~\cite{lutchyn2018majorana, aasen2016milestones}.

No wonder that the study of both equilibrium and nonequilibrium spectral and transport properties of these systems with engineered
superconducting state has become recently one of the central research directions in condensed matter physics.
While dc properties of these structures have been investigated in numerous theoretical and experimental
works, the dynamic effects and, in particular, high frequency response remains an appealing problem
which definitely deserves deeper understanding.
Indeed, the limitations on the operating frequencies for above mentioned
proximized devices~\cite{khaymovich2017nonlocality, semenoff2007stretched,
tewari2008testable, fu2010electron, bolech2007observing, nilsson2008splitting}
can be solely given by the dynamic characteristics of their induced
superconducting ordering. The nonlinear dynamic effects are also known to
provide a new route to the fascinating physics of coherent modification of the
density of states etc. induced by the microwave
irradiation~\cite{semenov2016coherent, devyatov2019relaxation}

Clearly, as
typical quantum computing devices operate at the temperatures {\it well below} the gap of the
primary superconductor,
the study of the relaxation dynamics of the order parameter {\it close} to the critical temperature of the superconducting transition
is irrelevant for their description.
More adequate theory  can be obtained by considering the so-called coherent quantum-mechanical dynamics of the system, which neglects inelastic scattering processes.
In addition, in superconducting systems with the unconventional pairing the dynamic
response is known to provide important information about the order parameter structure~\cite{bittner2015leggett,krull2016coupling}
working as a spectroscopic tool~\cite{fauseweh2017higgs}.
The analogous method can provide an insight about the internal structure of the primary and induced Cooper pairing in superconducting hybrids.

\begin{figure}
 \includegraphics[width=\linewidth]{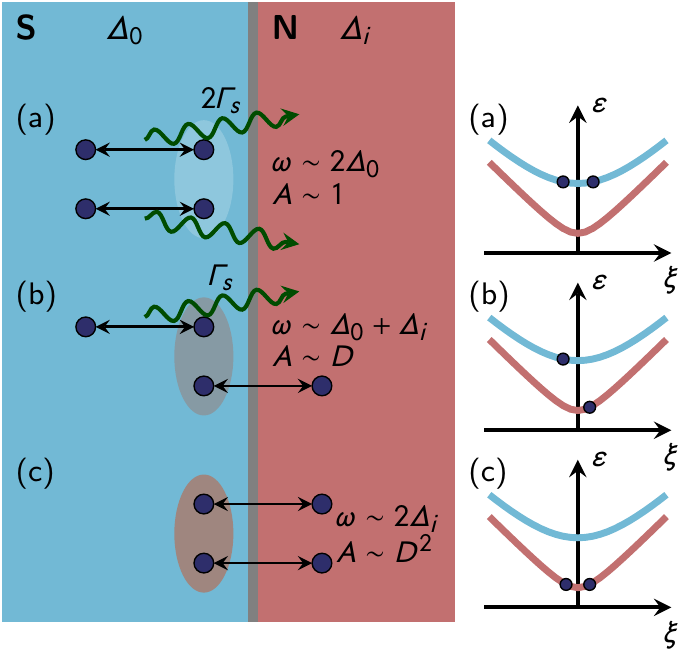}
 \caption{
  {\bf Illustration of three Higgs modes in SIN structure}, being coherent Cooper pair splitting-recovery processes with three different frequencies $\omega$, decay rates, and amplitudes $A$.
 }
 \label{fig:scheme}
\end{figure}

Indeed, even the linear dynamic response of the superconductor near equilibrium
provides a detection method of the superconducting gap structure via
the coherent order parameter oscillations known also as Higgs
modes~\cite{volkov1973collisionless-2, yuzbashyan2005integrable,
yuzbashyan2006relaxation, yuzbashyan2006dynamical,
yuzbashyan2006relaxation-err, higgs1964broken}.
The name is given
due to the analogy to the Higgs boson in particle
physics~\cite{higgs1964broken}.
In the low-temperature limit these near-equilibrium Higgs oscillations of the
the order parameter magnitude are described by the asymptotic long-time expression
$\Delta-\Delta_0 \sim \cos(2 \Delta_0t) / \sqrt{\Delta_0 t}$, where $\Delta_0$
is the superconducting gap in equilibrium~\cite{volkov1973collisionless-2, yuzbashyan2006relaxation,yuzbashyan2006relaxation-err,barankov2004collective, yuzbashyan2005integrable,
yuzbashyan2006dynamical}.
The Higgs mode has been first detected using the Raman spectroscopy in the
superconductors with charge density wave
ordering~\cite{sooryakumar1980raman, measson2015amplitude} as a peak at the
frequency $2\Delta_0$ in the Raman spectrum of
$2H$-$\mathrm{NbSe}_2$ below the superconducting transition temperature. Recent
progress of the THz experimental techniques allowed the direct observation of
the order parameter oscillations using the pump--probe
method~\cite{matsunaga2013higgs}. The broad band pump
excites a power--law relaxing Higgs mode at the frequency $2 \Delta_0$ while the
narrow band pump with a well defined frequency $\omega$ induces the oscillations
of the order parameter at the frequency $2 \omega$. The resonant third harmonic
generation~\cite{matsunaga2014light, silaev2019nonlinear} provides an evidence of the Higgs
mode excitation in the superconductors. An alternative method of the Higgs
mode detection through the second harmonic measurements has been recently for
the current carrying states~\cite{moor2017amplitude}.

In this Letter we address an important effect of the quasiparticle spectrum of the superconducting system on long-time dynamic properties of Higgs modes
and apply it to a system with proximity induced superconducting gap.
Key physical phenomena  related to the presence of the induced gap  are demonstrated on the example of a junction of the superconductor (S) and normal metal (N) coupled via
an insulating barrier (I) with a finite transparency.
We analyze the distinctive features of the
Higgs modes in this structure and make
the predictions for the experimentally accessible relevant quantities.

Let's first consider
the qualitative picture of the Higgs dynamics in SIN system, Fig.~\ref{fig:scheme},
before proceeding with the further microscopic calculations.
The Higgs modes in superconductors can be interpreted as a coherent splitting-recovery process
of Cooper pairs. The energy difference between the
ground state without quasiparticles and the excited state with two quasiparticles governs
the frequency $2 \Delta_0$ of this coherent superposition of above states as each unpaired quasiparticle
brings an additional energy $\Delta_0$.
Due to this simple qualitative reasoning we
may expect the frequencies of the Higgs modes to be
determined by the quasiparticle spectrum of the system.
{In SIN-structure t}he superconducting correlations penetrate {to} the normal metal
 and induce the hard gap $\Delta_i$ in the whole system~\cite{mcmillan1968tunneling}.
$\Delta_i$ depends on the transparency of the barrier and the size of the normal subsystem.

We claim that there are three Higgs modes in the SIN-structure corresponding to three
possible processes shown in Fig.~\ref{fig:scheme}.
First, as in the isolated superconductor the Cooper pair may coherently split
into two electrons both being located in the superconductor{,
Fig.~\ref{fig:scheme}(a)}.
The energy
$2 \Delta_0$ of the these unpaired electrons determines the frequency of this process.
However, in the presence of the normal metal the Cooper pair splitting can be accompanied by the
coherent tunneling process of either or both electrons, see Fig.~\ref{fig:scheme}(b) and (c), respectively.
The minimal energy of each electron which tunnels to the normal metal should be $\Delta_i$,
so the frequency of the corresponding Higgs mode is given by $\Delta_0 +
\Delta_i$ and $2\Delta_i$ for $n_e=1$ and $2$ electrons tunneling to the
normal metal, respectively.
The amplitudes $A$ of these modes are expected to be reduced by the factor of
$D^{n_e}$, where $D$ is transparency of the barrier (provided $D \ll 1$).
On top of that in the first two processes, Fig.~\ref{fig:scheme}(a) and (b), the coherent superposition can be
destroyed by the incoherent decay of each electron
located in the superconductor to the normal metal (see green wavy lines in Fig.~\ref{fig:scheme}) because the hard gap in the
spectrum of the whole system $\Delta_i$ is below the energy of quasiparticle in
the superconductor $\Delta_0$.
This effect results in an exponential damping of the Higgs modes with rate $\Gamma_s$ to each
$\Delta_0$ frequency contribution. The value $\Gamma_s$ is an
inverse lifetime of the electron in the superconductor  determined by the tunneling
rate from the superconductor to the normal metal.
To sum up, the main result of our work can be written from the qualitative perspective as
the following structure of the gap oscillations in SIN system:
\begin{multline}
\delta \Delta \sim
\frac{\cos(2 \Delta_0 t)}{\sqrt{\Delta_0 t}}e^{-2 \Gamma_s t} + \\
D \frac{\cos[(\Delta_0+\Delta_i) t]}{(\Delta_0 t)^p} e^{-\Gamma_s t} +
D^2 \frac{\cos(2\Delta_i t)}{(\Delta_0 t)^q} \ ,
\end{multline}
with certain power-law decay rates $p$ and $q$.

This qualitative picture can be confirmed by the direct microscopic calculations.
The considered SIN system can be described by the following Hamiltonian:
\begin{multline}
 \hat H = \sum\limits_{k\sigma} \xi_k^s \hat a_{k\sigma}^\dag \hat
 a_{k\sigma} + \sum\limits_k \left(\Delta \hat a_{k\uparrow}^\dag \hat a_{\bar k
 \downarrow}^\dag + \Delta^\ast \hat a_{\bar k \downarrow} \hat a_{k
 \uparrow}\right) + \\ + \sum\limits_{l\sigma} \xi_l^n \hat b_{l\sigma}^\dag
 \hat b_{l\sigma} + \sum\limits_{kl\sigma} \left(
 \gamma_{kl} \hat a_{k\sigma}^\dag \hat b_{l \sigma} + \gamma_{kl}^\ast
 \hat b_{l\sigma}^\dag \hat a_{k\sigma}
 \right)
\end{multline}
where $\hat a_{k \sigma}$ and $\hat a_{k\sigma}^\dag$ are the electron annihilation and
creation operators in the superconducting layer, $k$ is the index of the
single--electron state and $\sigma$ is the projection of the electron spin,
$\bar k$ denotes the index of the state obtained from the state $k$ by the
time inversion operation. The operators
$\hat b_{l \sigma}$ and $\hat b_{l \sigma}^\dag$ are the electron annihilation
and creation operators in the normal layer.
The last term describes the
tunneling between the superconductor and the normal metal.
Assuming the insulating interlayer to be dirty so that the electron momentum is not
conserved
we consider the tunneling matrix elements $\gamma_{kl}$ to be the Gaussian uncorrelated random values
$\langle \gamma_{kl} \gamma_{k'l'}^\ast \rangle = \gamma^2 \delta_{kk'}
\delta_{ll'}$~\cite{mcmillan1968tunneling,S-induced-footnote}.
This model approximately describes tunneling junction if the
magnitude of the tunneling matrix element $\gamma^2$ is proportional to $S /
(V_s V_n)$, where $S$ is the junction area, $V_s$ and $V_n$ are the volumes of
the superconducting and normal subsystems, respectively.
The superconducting order parameter $\Delta$ is given by the self-consistency
equation:
\begin{equation}
 \Delta = \frac{\lambda}{V_s} \sum\limits_k \langle \hat a_{k\uparrow} \hat a_{\bar k
 \downarrow}\rangle \ ,
 \label{eq:selfconsistency}
\end{equation}
where $\lambda$ is the pairing constant and $V_s$ is the volume of the
superconductor, the brackets $\langle \ldots \rangle$ denote a quantum-mechanical averaging.
 Here we assume the thickness of the superconducting subsystem to be small
compared to the superconducting coherence length so we can consider $\Delta$ to be homogeneous
within
the sample.

\begin{figure}[t]
 \includegraphics[width=\linewidth]{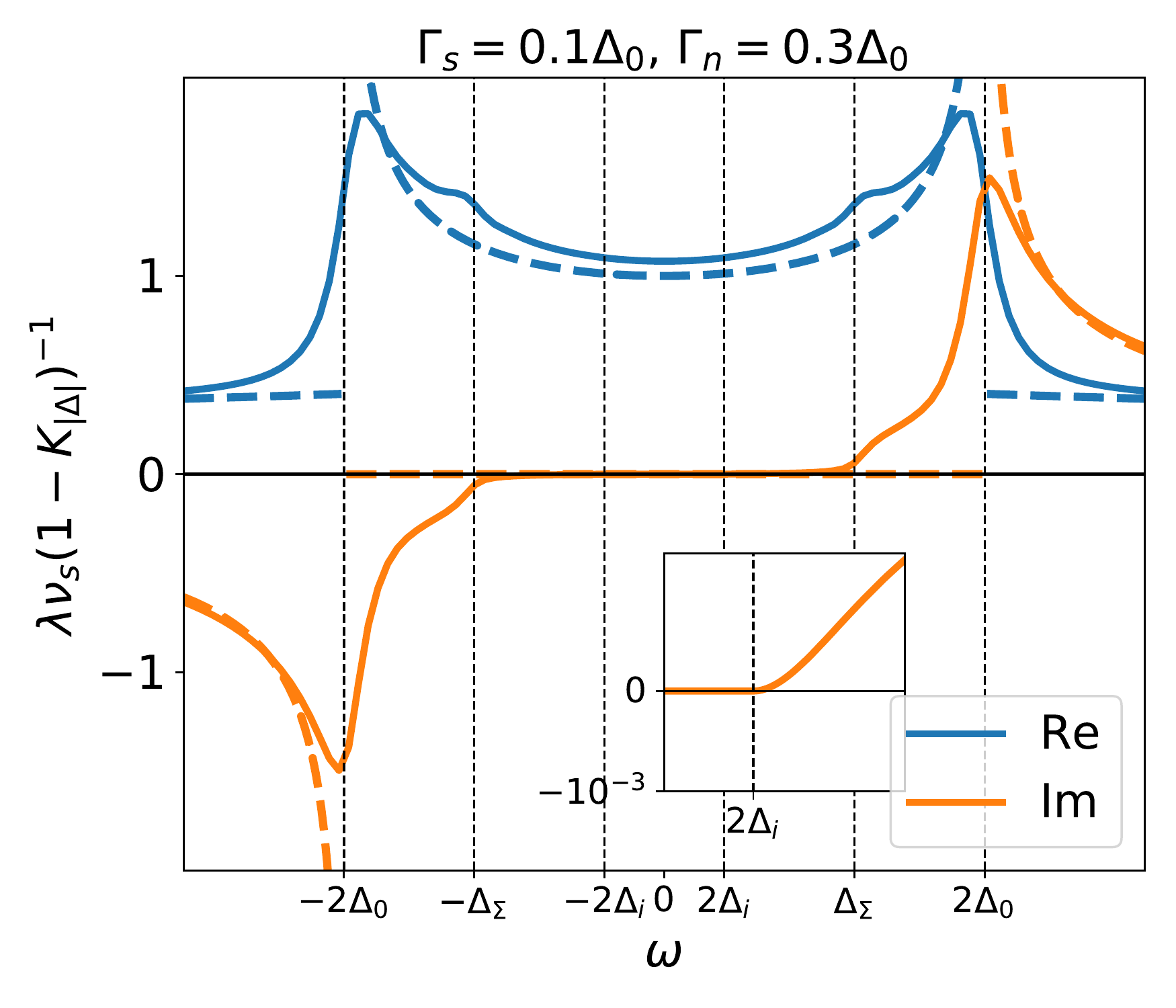}
 \caption{The real and imaginary part of $[1 - K_\Delta(\omega)]^{-1}$ in
 the case of the finite tunneling rates $\Gamma_s$ and $\Gamma_n$. The
inset shows the singularity at the frequency $2 \Delta_i$ which cannot be
observed in the regular scale. For the given parameters $\Gamma_s = 0.1
\Delta_0$ and $\Gamma_n = 0.3 \Delta_0$ the induced gap $\Delta_i$ is
approximately equal to $0.2 \Delta_0$. Here $\Delta_\Sigma = \Delta_0 +
\Delta_i$. The dashed line shows $[1 - K_\Delta(\omega)]^{-1}$ for the
isolated superconductor $\Gamma_s = \Gamma_n = 0$.}
 \label{fig:kernel}
\end{figure}

We study the dynamics of the system using non-equilibrium Keldysh technique.
Following approach developed in the Refs.~\cite{mcmillan1968tunneling,
volkov1995proximity, kopnin2011proximity, kopnin2013predicted, kopnin2013vortex} we treat the tunneling operator
within
the self-consistent Born approximation
 and write the following set of the
equations for the Green's functions in the superconductor and normal metal:
\begin{equation}
 \begin{gathered}
 i \frac{\partial \breve G_k^{s}}{\partial t} - \breve H_k^s
 \breve{G}_{k}^{s} - \breve \Sigma^s \ast \breve G_k^{s} =
 \delta(t-t') \ , \\
 i \frac{\partial \breve G_l^{n}}{\partial t} - \breve H_l^n
 \breve{G}_{l}^{n} - \breve \Sigma^n \ast \breve G_l^{n} =
 \delta(t-t') \ ,
 \end{gathered}
 \label{eq:dyson}
\end{equation}
where the Green's function of the superconductor $\breve G_k^{s}$ and the
Green's function of the normal metal $\breve G_l^n$ are the $4 \times 4$ matrices in the Keldysh--Nambu space
\begin{equation}
 \breve G_{k/l}^{s/n} = \begin{pmatrix}
 \check G_{k/l}^{s/n(R)} & \check G_{k/l}^{s/n(K)} \\
 0 & \check G_{k/l}^{s/n(A)}
 \end{pmatrix} \ ,
\end{equation}
and $\breve H_k^s$ and $\breve H_l^n$ are the single mode Hamiltonians of the
superconductor and the normal metal:
\begin{gather}
    \breve  H_{k/l}^{s/n} = \begin{pmatrix}
        \check H_{k/l}^{s/n} & 0 \\
        0 & \check H_{k/l}^{s/n}
    \end{pmatrix}, \\
    \check H_{k}^s = \begin{pmatrix}
        \xi_{k}^s & \Delta \\
        \Delta^\ast & -\xi_k^s
    \end{pmatrix},~
    \check H_{l}^n = \begin{pmatrix}
        \xi_{l}^n & 0 \\
        0 & -\xi_l^n
    \end{pmatrix} \ .
\end{gather}
The self-energies of the superconductor $\breve \Sigma^s$ and the normal metal
$\breve \Sigma^n$ describe
the tunneling between two subsystems and self-consistently expressed through the Green's
functions $\breve G_l^n$ and $\breve G_k^s$, respectively.
These self-energies are proportional to the tunneling rates of electrons from
the superconductor
$\breve \Sigma_s \propto \Gamma_s = \pi \gamma^2 \nu_n V_n$ and the normal
metal $\breve \Sigma_n \propto \Gamma_n = \pi \gamma^2 \nu_s
V_s$, where $\nu_s$ and $\nu_n$ are the densities of normal states per the unit volume in the
superconductor and the normal metal at the Fermi level, respectively. The
derivation of the Dyson equations~(\ref{eq:dyson}) and the explicit form of the self-energies
are given in Appendix. 

In order to study the near--equilibrium dynamics of the system we expand the order
parameter near $\Delta_0$ assuming its phase to be zero in equilibrium without loss of the generality:
\begin{equation}
 \Delta(t) = \Delta_0 + \delta \Delta(t) + i \Delta_0 \delta \theta(t) \ .
\end{equation}
Here $\delta \Delta$ and $\delta \theta$ are the perturbations of the magnitude and the phase of $\Delta$, respectively.
One can introduce
the corresponding perturbations to the Green's functions
and the self-energies and linearize the system of the equations for the Green's functions
with respect to the perturbation of the superconducting order parameter.
Along with the linearized self-consistency
equation~(\ref{eq:selfconsistency}) this system can be reduced to the equations for the
eigenmodes of the order parameter in the Fourier form (see Appendix for details):
\begin{equation}
 \begin{gathered}
 \delta \Delta (\omega) = K_{\Delta}(\omega) \delta
 \Delta(\omega) + K'(\omega) \Delta_0 \delta \theta(\omega) \\
 \delta \theta (\omega) = K''(\omega) \frac{\delta
 \Delta(\omega)}{\Delta_0} + K_\theta(\omega) \delta \theta(\omega) \\
 \end{gathered}
 \label{eq:higgs_goldstone_modes}
\end{equation}
The off-diagonal kernels $K'$ and $K''$ are equal exactly to zero in the systems which
have electron--hole symmetry and can be neglected if the Fermi level both in
the superconductor and the normal metal is far from the van Hove singularities
in the density of states~\cite{vonHove-footnote}.
The singular points of $[1-K_{\Delta}(\omega)]^{-1}$ and
$[1-K_{\theta}(\omega)]^{-1}$ correspond to the frequencies of the Higgs and
Anderson-Bogoliubov modes of the superconductor.
The perturbations of the magnitude and the phase of the order
parameter are completely independent so hereafter we focus only on the study of
the Higgs modes.

The linear response of the superconducting order parameter to an external
force $f(\omega)$ takes the following form:
\begin{equation}
 \delta \Delta(t) = \frac{1}{2\pi} \int \frac{f(\omega) e^{-i \omega
 t}\;d\omega}{1 - K_\Delta(\omega)} \ .
 \label{eq:higgs}
\end{equation}
Previous studies \cite{tsuji2015theory,matsunaga2013higgs,matsunaga2014light,
silaev2019nonlinear}
show that this force can originate, e.g.,
from the pulses of the external electromagnetic
field. In Fig.~\ref{fig:kernel} the typical frequency dependencies of the real
and imaginary parts of $[1-K_\Delta(\omega)]^{-1}$ are shown for some
particular values of the tunneling rates.
The spectrum of the Higgs modes appears to
be consistent with the picture shown in Fig.~\ref{fig:scheme}.
The broadened features at the frequencies
$\omega \approx \pm 2 \Delta_0$ and $\omega \approx \pm (\Delta_0 + \Delta_i)$
are seen clearly, while the singularity at $\omega = \pm 2 \Delta_i$ can be seen
only in the zoomed inset. The latter singularity corresponds to the low frequency
Higgs mode, Fig.~\ref{fig:scheme}(c). This mode has no exponential damping as $\Delta_i$ is a hard gap of the whole system (the singular point of the kernel is
exactly at the real axis). It means that this low-frequency mode gives the
major contribution to the oscillations of the order parameter in the long time
limit, $t\gg \Gamma_s^{-1}, \Gamma_n^{-1}$, however the amplitude of this mode
is few orders of magnitude lower than the amplitude of
the usual Higgs mode due to the low transparency of the insulating barrier.

The kernel $K_{\Delta}$ can be evaluated analytically in the zero
temperature limit and $\Gamma_n
= 0$ which corresponds to the bulk normal metal $V_n \to \infty$  with the
suppressed induced superconducting ordering and vanishing induced gap
$\Delta_i = 0$:
\begin{multline}
    \frac{1 - K_\Delta(\omega)}{\lambda \nu_s} =
       i \frac{\sqrt{4 \Delta_0^2 - \omega_2^2}}{2 \omega_2} \cdot \ln \left[
       \frac{F(\omega_1, \omega_2)}{F(i \Gamma_s, \omega_2)}\right]+
        \\
   i \frac{ \sqrt{4 \Delta_0^2 - { \omega^2}}}{2  \omega}
        \ln \left[
            \frac{
                F(-i \Gamma_s, \omega)
            }{
                F(\omega_1, \omega)
            } \cdot
            \frac{\omega + i \sqrt{4 \Delta_0^2 -  \omega^2}}
            { -\omega + i \sqrt{4 \Delta_0^2 -  \omega^2}}
        \right]
    \ ,
\end{multline}
where $\omega_1 = \omega + i \Gamma_s$, $\omega_2 = \omega + 2 i \Gamma_s$,
and $F(x, y) = 2 \Delta_0^2 - x y + \sqrt{\Delta_0^2 - x^2} \sqrt{4 \Delta_0^2 - y^2}$.
%
 In the isolated superconductor, $\Gamma_s = 0$, the kernel as a
function of the complex frequency $\omega$ has two
branch points at $\omega = \pm 2 \Delta_0$ which give the usual polynomially
damped Higgs mode in the superconductor.
In the presence of tunneling, $\Gamma_s>0$, these branch points shift to the points $\omega = \pm
2 \Delta_0 - 2 i \Gamma_s$  corresponding to exponential damping of this mode.
 Moreover, two additional branch points at $\omega = \pm\Delta_0 - i \Gamma_s$
 appear triggering a new Higgs mode, which has already revealed itself in
 Fig.~\ref{fig:kernel}.   The absence of the low frequency mode at $2 \Delta_i$ is  rather expected for the
considered case $\Gamma_n = 0$ because the electrons of the normal metal
are not affected by the proximity with the superconductor  and, thus, cannot form a Cooper pair as in
Fig.~\ref{fig:scheme}(c). As a result they do not contribute to the order parameter oscillations.
The asymptotic behavior of the order parameter perturbation in the intermediate-time limit $\Delta_0^{-1} \ll t \ll \Gamma_s^{-1}$ reads as
\begin{multline}\label{eq:higgs_kernel}
 \delta \Delta(t) \approx -i \sum\limits_j
 \frac{e^{-i \omega_j t} \mathop{\mathrm{res}}_{\omega_j} f(\omega)}{1 -
 K_\Delta(\omega_j)} +\\ + \frac{a \cos(2 \Delta_0 t - \alpha)}{\sqrt{\Delta_0 t}}
 e^{-2 \Gamma_s t} + \frac{\Gamma_s}{\Delta_0} \frac{b \cos(\Delta_0 t
 - \beta)}{(\Delta_0 t)^{3/2}} e^{-\Gamma_s t} \ .
\end{multline}
The details of calculations, restrictions on the analytic properties of the exciting force
$f(\omega)$, and  the expressions of parameters $a$, $b$, $\alpha$ and $\beta$ depending on the
exciting force are given in Appendix.
The first term in~\eqref{eq:higgs_kernel} describes the forced oscillations of the order parameter which
occur at the frequencies corresponding to the poles
of the Fourier spectrum of the external force $f(\omega)$. The finite tunneling rate $\Gamma_s$ leads to the exponential damping of the oscillations of the order parameter and appearance of a new Higgs mode at
the frequency $\Delta_0$ with the magnitude suppressed by factor $\Gamma_s /
\Delta_0$. This mode corresponds to the middle-frequency mode $\Delta_0 +
\Delta_i$ shown in Fig.~\ref{fig:scheme}(b) in the limit $\Delta_i = 0$. Let's
emphasize once again that all the above analysis was related to the case of
the small tunneling rates $\Gamma_s \ll \Delta_0$. In the opposite limit
$\Gamma_s \gg \Delta_0$ we get
the gapless superconductivity with a pure imaginary frequency of the Higgs
mode: $\omega = -i \Delta_0^2 / \Gamma_s$. The relaxation of the
order parameter in this limit is  described by the decaying exponent
 $\delta \Delta(t) \propto \exp(-\Delta_0^2 t / \Gamma_s)$.

The Higgs modes of the SIN system can be studied using a pump--probe experiment similar to the
one developed in the Ref.~\cite{matsunaga2013higgs}. The measured $\delta
\Delta(t)$ can be analyzed using the Fourier transform. The Fourier spectrum is
expected to have two peaks at $\omega \approx 2 \Delta_0$ and $\omega \approx
\Delta_0 + \Delta_i$  corresponding to  the above Higgs modes of the
SIN system.
In the limit of low transparency $D\ll 1$ the mode with the frequency $2 \Delta_i$ has too low magnitude so
it's experimental observation can be hampered,
 however it may be visible at intermediate transparencies $D\sim 1$, $\Delta_i\lesssim \Delta_0$.
Another way to detect the
Higgs modes is the experimental studying of the frequency dependence
of the third harmonic generation~\cite{silaev2019nonlinear,matsunaga2014light}. The electromagnetic wave with the frequency
$\Omega$ excites the Higgs mode with the frequency $2 \Omega$. The magnitude
of the generated nonlinear signal with the frequency $3 \Omega$ should depend on the magnitude
of the oscillations of the order parameter and therefore one can expect the appearance of the broadened
resonances in the third harmonic response if the frequency $2 \Omega$ is
 close to the frequency of any of the
Higgs modes.
However,
the effect of generation of non-equilibrium quasiparticles by
the electromagnetic radiation with the frequency $\Omega>2 \Delta_i$ may complicate the observation of the resonant effect in
the third harmonic generation. The overheating  can be significantly reduced
at intermediate transparencies $D\sim 1$ when the induced gap is high enough
$3 \Delta_i > \Delta_0$.  This improves the observability of the resonance at
$2 \Omega = \Delta_0 + \Delta_i$ predicted above.

To sum up, the effect of the quasiparticle spectrum of the superconducting system on
the low-temperature dynamics of the order parameter has been studied on an example of a hybrid superconductor-insulator-normal metal system.
 Two new Higgs modes in such system have been discovered. The
frequencies of these modes are formed by sums of the two quasiparticle
energies, which are in a good correspondence with the qualitative
interpretation of the Higgs modes as of coherent processes of splitting and
recovery of the Cooper pairs, Fig.~\ref{fig:scheme}.
The proposals of experimental observation of these new Higgs modes in hybrid SIN system have been developed
 basing on the  existing THz techniques used for study of
the Higgs modes in the pure superconductors.


This work was supported, in part, by
the Russian Foundation for Basic Research under Grant No. 17-52-12044,
the Foundation for the Advancement to Theoretical Physics and Mathematics
``BASIS'' Grant No. 17-11-109, and
the German-Russian Interdisciplinary Science Center (G-RISC) funded by the German Federal
Foreign Office via the German Academic Exchange Service (DAAD).
In the part of numerical calculations,
the work was supported by Russian Science Foundation (Grant No. 17-12-01383).
I. M. K. acknowledges the support of the German Research Foundation (DFG) Grant No. KH~425/1-1.
V. L. V. and A. S. M. appreciate warm hospitality of the Max-Planck Institute for the Physics
of Complex Systems, Dresden, Germany, extended to them during their visits when this work was done.

%


\begin{widetext}

\appendix
\section{Derivation of the Dyson equations}

In this section we derive the equations for the Green's functions of the
superconductor and the normal metal.
We define the Nambu pseudospinor operators as follows:
\begin{equation}
    \hat A_k = \begin{pmatrix}
        \hat a_{k\uparrow} \\
        \hat a_{\bar k \downarrow}^\dag
    \end{pmatrix},\quad
    \hat B_l = \begin{pmatrix}
        \hat b_{l\uparrow} \\
        \hat b_{\bar l \downarrow}^\dag
    \end{pmatrix} \ .
\end{equation}
Using these operators we define the retarded, advanced and Keldysh Green's
functions of the superconductor in the following way:
\begin{equation}
    \begin{gathered}
        \check{ G}_{kk'}^{ss(R)}(t,t') = -i \Theta(t - t') \left\langle \hat A_k(t) \hat
        A_{k'}^\dag(t') + \hat A_{k'}^{\dag T} (t') \hat A_k^T(t)\right
        \rangle \\
        \check{ G}_{kk'}^{ss(A)}(t,t') = i \Theta(t' - t) \left\langle \hat A_k(t) \hat
        A_{k'}^\dag(t') + \hat A_{k'}^{\dag T} (t') \hat A_k^T(t)\right
        \rangle \\
        \check{ G}_{kk'}^{ss(K)}(t,t') = -i \left\langle
        \hat A_k(t) \hat A_{k'}^\dag(t') - \hat A_{k'}^{\dag T}(t') \hat
        A_{k}^T(t)
        \right \rangle
    \end{gathered}
\end{equation}
The Green's functions of the normal metal $G_{ll'}^{ll(RAK)}$ and the
tunneling Green's functions $G_{kl}^{sn(RAK)}$ and $G_{lk}^{ns(RAK)}$ are
defined in the same way replacing the operators $\hat A$ with $\hat B$.
One can construct $4 \times 4$ matrix Green's functions in the Keldysh--Nambu
space in the usual way:
\begin{equation}
    \breve{ G}^{\alpha \beta}_{kk'}(t,t') = \begin{pmatrix}
        \check{ G}_{kk'}^{\alpha\beta (R)}(t,t') &
        \check{ G}_{kk'}^{\alpha \beta (K)}(t,t') \\
        0 & \check{ G}_{kk'}^{\alpha \beta (A)}(t,t')
    \end{pmatrix} \ ,
\end{equation}
where the indices $\alpha$ and $\beta$ denote $s$ and $n$. The equations for
the Green's functions read as:
\begin{equation}
    \begin{gathered}
        i \frac{\partial}{\partial t} \breve{ G}_{kk'}^{ss} - \breve
        H_k^s \breve{ G}_{kk'}^{ss} - \sum\limits_l \gamma_{kl} \breve
        \tau_3 \breve{ G}_{lk'}^{ns} =  \delta(t - t')
        \delta_{kk'} \\
        i \frac{\partial}{\partial t} \breve{ G}_{lk}^{ns} - \breve
        H_l^n \breve{ G}_{lk}^{ns} - \sum\limits_{k'}
        \gamma_{k'l}^\ast \breve
        \tau_3 \breve{ G}_{k'k}^{ss} = 0
    \end{gathered} \ ,
\end{equation}
\begin{equation}
    \begin{gathered}
        i \frac{\partial}{\partial t} \breve{ G}_{ll'}^{nn} - \breve
        H_n^s \breve{ G}_{ll'}^{nn} - \sum\limits_k \gamma_{kl}^\ast \breve
        \tau_3 \breve{ G}_{kl'}^{sn} =  \delta(t - t')
        \delta_{ll'} \\
        i \frac{\partial}{\partial t} \breve{ G}_{kl}^{sn} - \breve
        H_k^s \breve{ G}_{kl}^{sn} - \sum\limits_{l'}
        \gamma_{kl'} \breve
        \tau_3 \breve{ G}_{l'l}^{nn} = 0
    \end{gathered} \ ,
\end{equation}
where
\begin{equation}
    \breve H_k^s = \begin{pmatrix}
        \check H_k^s & 0 \\
        0 & \check H_k^s
    \end{pmatrix},~
    \check H_k^s = \begin{pmatrix}
        \xi_k^s & \Delta \\
        \Delta^\ast & -\xi_k^s
    \end{pmatrix} \ ,
\end{equation}
\begin{equation}
    \breve H_l^n = \begin{pmatrix}
        \check H_l^n & 0 \\
        0 & \check H_l^n
    \end{pmatrix},~
    \check H_l^n = \begin{pmatrix}
        \xi_l^n & 0 \\
        0 & -\xi_l^n
    \end{pmatrix} \ ,
\end{equation}
\begin{equation}
    \breve \tau_3 = \begin{pmatrix}
        \check \tau_3 & 0 \\
        0 & \check \tau_3
    \end{pmatrix},~
    \check \tau_3 = \begin{pmatrix}
        1 & 0 \\
        0 & -1
    \end{pmatrix} \ .
\end{equation}
The selfconsistency condition reads as
\begin{equation}
    \Delta(t) = \frac{i \lambda}{4 V} \sum\limits_k \Tr \left[
        \left(\check \tau_1 + i \check \tau_2\right) \check { G}_{kk}^{ss(K)}(t,
        t)
    \right] \ ,
    \label{eq:selfconsistency}
\end{equation}
where
\begin{equation}
    \check \tau_1 = \begin{pmatrix}
        0 & 1 \\
        1 & 0
    \end{pmatrix},\quad
    \check \tau_2 = \begin{pmatrix}
        0 & -i \\
        i & 0
    \end{pmatrix} \ .
\end{equation}
Using the Green's functions of the isolated superconductor and the normal
metal $\breve{\mathcal G}_{k}^s$ and $\breve{\mathcal G}_{l}^n$ we may
eliminate the tunneling Green's functions:
\begin{gather}
    \breve{ G}_{lk}^{ns} = \sum\limits_{k'} \gamma_{k'l}^\ast
    \breve{\mathcal G}_{l}^{n}  \breve \tau_3 \ast \breve{
    G}_{k'k}^{ss} \ , \\
    \breve{ G}_{kl}^{sn} = \sum\limits_{l'} \gamma_{kl'}
    \breve{\mathcal G}_{k}^{s}  \breve \tau_3 \ast \breve{
    G}_{l'l}^{nn} \ ,
\end{gather}
where $\ast$ denotes the convolution operation
\begin{equation}
    (X \ast Y)(t,t') = \int X(t,t'') Y(t'', t')\;dt''
\end{equation}
and the Green's functions $\breve{\mathcal G}_{k}^s$ and $\breve{\mathcal
G}_{l}^n$ satisfy the following equations:
\begin{gather}
    i\frac{\partial}{\partial t} \breve{\mathcal G}_k^s - \breve H_k^s
    \breve{\mathcal G}_k^s =
     \delta(t-t') \ ,\\
    i \frac{\partial}{\partial t} \breve{\mathcal G}_l^n - \breve H_l^n
    \breve{\mathcal G}_l^n =
     \delta(t - t') \ .
\end{gather}
Thus, we can write two independent equations for the Green's functions in
the superconductor and the normal metal:
\begin{equation}
    i \frac{\partial}{\partial t}\breve{ G}_{kk'}^{ss} - \breve H_k^s
    \breve{ G}_{kk'}^{ss} -  \sum\limits_{k''} \breve \Sigma^s_{kk''}  \ast
    \breve{ G}_{k''k'}^{ss} =  \delta(t-t') \delta_{kk'} \ ,
\end{equation}
\begin{equation}
    i\frac{\partial}{\partial t}\breve{ G}_{ll'}^{nn} - \breve H_l^n
    \breve{ G}_{ll'}^{nn} -
    \sum\limits_{l''}  \breve \Sigma^n_{ll''}\ast \breve{ G}_{l''l'}^{nn} =
    \delta(t-t') \delta_{ll'} \ ,
\end{equation}
where the self-energies of the superconductor and the normal metal are:
\begin{equation}
    \breve \Sigma^s_{kk''} = \sum \limits_l \gamma_{kl}
    \gamma_{k''l}^\ast \breve \tau_3 \breve{\mathcal G}_l^n \breve \tau_3 \ ,
\end{equation}
\begin{equation}
    \breve \Sigma^n_{ll''} = \sum \limits_k \gamma_{kl}^\ast \gamma_{kl''} \breve \tau_3 \breve
    {\mathcal G}_{k}^s \breve \tau_3 \ .
\end{equation}

\begin{figure*}
    \subfigure[exact Dyson equation for the Green's functions (before
    averaging)]{
        \includegraphics[width=0.31\linewidth]{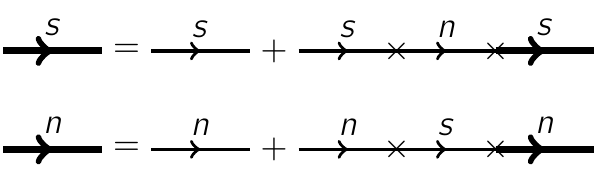}
    }
    \subfigure[selfconsistent Born approximation for the averaged Green's functions]{
        \includegraphics[width=0.31\linewidth]{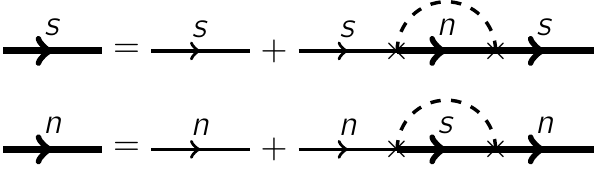}
    }
    \caption{Diagrams for the Green's functions in the superconductor and the
        normal metal. The symbol $\times$ denotes tunneling, the dashed line
        denotes the correlator between the matrix elements of the tunneling
    operator.}
    \label{fig:diagram}
\end{figure*}

These equations written in the diagram form are shown in the
Fig.~\ref{fig:diagram}(a). These equations are not practical to use as they
contain the tunneling matrix elements $\gamma_{kl}$ which are the random
numbers. One can average the equations over the random matrix elements, the
average of the product of the matrix elements can be expanded
as sums of correlators due to the Wick theorem, thus, the averaged Green's
function can be written as a sum of diagrams. We omit the diagrams with
the intersecting correlators (thus, we neglect the vertex corrections) and
take account only of the diagrams with the consequent and nested correlators.
Such approach is known as selfconsistent Born approximation. The diagrammatic
form of the Dyson equation for this approximation is shown in the
Fig.~\ref{fig:diagram}(b). After the averaging the self-energies and the Green's function appear
to be diagonal in the normal mode picture and obey the following equations:

\begin{equation}
    \begin{gathered}
        i \frac{\partial \breve G_k^{s}}{\partial t} - \breve H_k^s
        \breve{G}_{k}^{s} - \breve \Sigma^s \ast \breve G_k^{s} =
        \delta(t-t') \ , \\
        i \frac{\partial \breve G_l^{n}}{\partial t} - \breve H_l^n
        \breve{G}_{l}^{n} - \breve \Sigma^n \ast \breve G_l^{n} =
        \delta(t-t') \ ,
    \end{gathered}
    \label{eq:dyson}
\end{equation}

\begin{equation}
    \begin{gathered}
        \breve \Sigma^s = \gamma^2 \sum\limits_l \breve \tau_3 \breve G_l^n
        \breve \tau_3 \ , \\
        \breve \Sigma^n = \gamma^2 \sum\limits_k \breve \tau_3 \breve G_k^s
        \breve \tau_3 \ .
    \end{gathered}
\end{equation}
In the wide band approximation the sum over the normal modes can be replaced
with the integral over the normal energy $\sum_k \to \nu_s V_s \int
\;d\xi_k^s$ ($\sum_l \to \nu_n V_n \int
\;d\xi_l^n$):

\begin{equation}
    \begin{gathered}
        \breve \Sigma^s = \frac{\Gamma_s}{\pi} \int \breve \tau_3 \breve G_l^n \breve
        \tau_3\;d\xi_l^n\ , \\
        \breve \Sigma^n =  \frac{\Gamma_n}{\pi} \int \breve \tau_3 \breve G_k^s \breve
        \tau_3\;d\xi_k^s \ .
    \end{gathered}
\end{equation}
Here we have introduced the tunneling rates $\Gamma_s = \pi \gamma^2 \nu_n
V_n$ and $\Gamma_n = \pi \gamma^2 \nu_s V_s$.

\section{Derivation of the equations for the eigenmodes}
Let us introduce the perturbations of the Green's functions and the
self-energies with respect to the perturbation of the magnitude $\delta
\Delta$ and the phase $\delta \theta$
of the superconducting order
parameter:
\begin{equation}
    \begin{gathered}
        \breve G_k^{s} = \breve G_{0k}^{s} + \delta \breve G_{k}^{s},~\breve
        G_k^{n} = \breve G_{0l}^{n} + \delta \breve G_{l}^{n} \ ,\\
        \breve \Sigma^{s/n} = \breve \Sigma_{0}^{s/n} + \delta \breve
        \Sigma^{s/n} \ ,
    \end{gathered}
\end{equation}
where $\breve G_{0k}^s$, $\breve G_{0l}^n$ and $\breve \Sigma^{s/n}$ are the
equilibrium Green's
functions and self-energies of the superconductor and the normal metal with
the account of tunneling.
The closed set of equations for the linear perturbations of the Green's
functions and the self-energies reads as:
\begin{equation}
    \begin{gathered}
        i\frac{\partial}{\partial t} \delta \breve G_k^s - H_{0k}^s \delta G_k^s - \Sigma^s \ast
        \delta \breve G_k^s - \delta \Sigma^s \ast \breve G_{0k}^s = \delta
        H^s G_{0k}^s \\
        i\frac{\partial}{\partial t} \delta \breve G_l^n - H_{l}^n \delta G_l^n - \Sigma^n \ast
        \delta \breve G_l^n - \delta \Sigma^n \ast \breve G_{0l}^n = 0 \\
        \delta \breve \Sigma^s =  \frac{\Gamma_s}{\pi} \int \breve \tau_3
        \delta \breve G_l^n \breve
        \tau_3\;d\xi_l^n,\\
        \delta \breve \Sigma^n =  \frac{\Gamma_n}{\pi} \int \breve \tau_3
        \delta \breve G_k^s \breve
        \tau_3\;d\xi_k^s \ .
    \end{gathered}
    \label{eq:dyson-linear}
\end{equation}
The perturbation to the single mode Hamiltonian of the superconductor is $\delta \breve H^s = \delta
\Delta \breve \tau_1 - i \Delta_0 \delta \theta \breve \tau_2$. One can easily
solve the equations for $\delta G$ in the Fourier form:
\begin{equation}
    \begin{gathered}
        \delta \breve G_k^s(\omega,\omega') = \breve G_{0k}^s(\omega) \left[
            \delta \breve H^s(\omega - \omega') + \delta \Sigma^s(\omega,
            \omega')
        \right] \breve G_{0k}^s(\omega') \ , \\
        \delta \breve G_l^n(\omega,\omega') = \breve G_{0l}^n(\omega)
            \delta \Sigma^n(\omega,
            \omega')
        \breve G_{0l}^n (\omega') \ ,
    \end{gathered}
\end{equation}
where the Fourier transform of the Green's functions is defined as follows:
\begin{equation}
    \begin{gathered}
        \delta \breve G(t, t') = \frac{1}{(2\pi)^2} \int \delta \breve G(\omega,
        \omega') e^{-i\omega t + i \omega' t'}\;d\omega\;d\omega' \ , \\
        \breve G_0(t-t') = \frac{1}{2\pi} \int \breve G_0(\omega) e^{-i\omega(t-t')}
        \;d\omega \ .
    \end{gathered}
\end{equation}
We introduce the quasiclassic Green's function $\delta \breve g^s = \int
\delta \breve G_k^s\;d\xi_k^s$ and write an algebraic equation for it:
\begin{multline}
    \delta \breve g(\omega, \omega') = \int \breve G_{0k}^s(\omega) \delta \breve
    H_k(\omega - \omega') \breve G_{0k}^s(\omega')\;d\xi_k^s +  \frac{\Gamma_n
    \Gamma_s}{\pi^2} \int \breve G_{0k}^s(\omega) \breve \tau_3 \breve
    G_{0l}^n(\omega) \breve \tau_3 \delta \breve g(\omega, \omega')  \breve \tau_3
    \breve G_{0l}^n(\omega') \breve \tau_3 \breve
    G_{0k}^s(\omega')\;d\xi_k^s\;d\xi_l^n
\end{multline}
The above
equation can be considered as a system of 12 linear equations for the 12
components of matrix Green's function $\breve g(\omega, \omega')$ (retarded,
advanced and Keldysh, each of them is $2\times 2$ matrix). Each integral can
be evaluated analytically because the equilibrium Green's functions are
rational functions of the normal energies $\xi_k^s$ and $\xi_l^n$. The solution can be
written in the following form:
\begin{equation}
    \delta \breve g(\omega, \omega') = \breve A_\Delta(\omega, \omega') \delta \Delta
    (\omega - \omega') + \Delta_0 \breve
    A_\theta (\omega, \omega') \delta \theta(\omega - \omega') \ .
\end{equation}
One should  use the selfconsistency equation~(\ref{eq:selfconsistency})
\begin{equation}
    \begin{gathered}
        \delta \Delta(\omega) = \frac{i \lambda \nu_s}{8 \pi } \int \Tr
        \check \tau_1 \delta \check g^{(K)}(\omega' + \omega, \omega')\;d\omega' \ , \\
        \delta \theta(\omega) = \frac{i \lambda \nu_s}{8 \pi  \Delta_0} \int \Tr
        \check \tau_2 \delta \check g^{(K)}(\omega' + \omega, \omega')\;d\omega'
    \end{gathered}
\end{equation}
and obtain the expression for all the kernels in the equation
(7) of the main text of the paper:
\begin{equation}
    \begin{gathered}
        K_\Delta(\omega) = \frac{i \lambda \nu_s}{8 \pi } \int \Tr \check
        \tau_1 \check A_\Delta^{(K)}(\omega' + \omega, \omega')\;d\omega' \ , \\
        K'(\omega) = \frac{i \lambda \nu_s}{8 \pi } \int \Tr \check
        \tau_1 \check A_\theta^{(K)}(\omega' + \omega, \omega')\;d\omega' \ , \\
        K''(\omega) = \frac{i \lambda \nu_s}{8 \pi } \int \Tr \check
        \tau_2 \check A_\Delta^{(K)}(\omega' + \omega, \omega')\;d\omega' \ ,
        \\
        K_\theta(\omega) = \frac{i \lambda \nu_s}{8 \pi } \int \Tr \check
        \tau_2 \check A_\theta^{(K)}(\omega' + \omega, \omega')\;d\omega' \ .
    \end{gathered}
\end{equation}

\section{Bulk normal metal $\Gamma_n = 0$}

\begin{figure*}
    \subfigure[ ]{
    \includegraphics[width=0.45\linewidth]{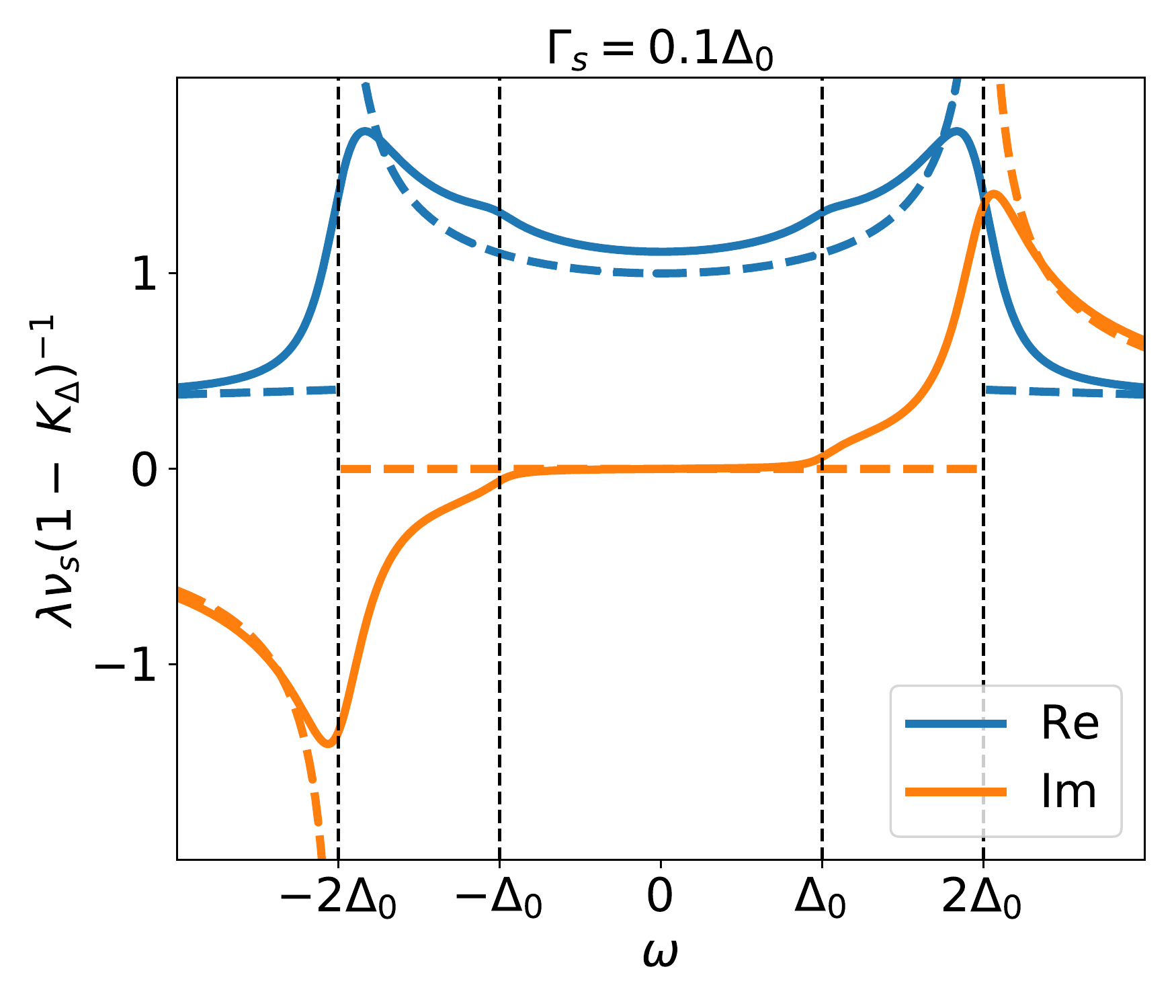}}
    \subfigure[ ]{
    \includegraphics[width=0.45\linewidth]{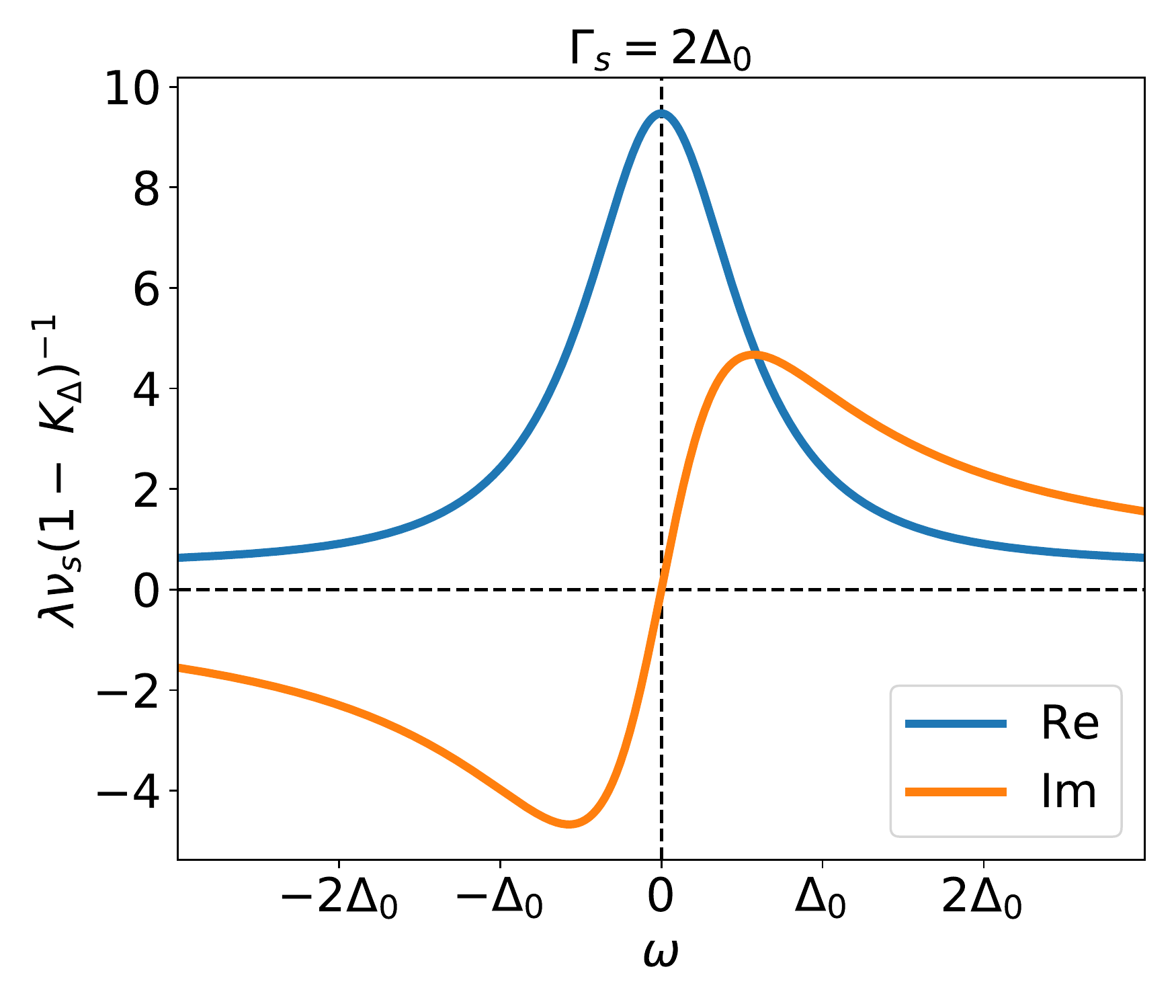}}
 \caption{The plots of $[1-K_\Delta(\omega)]^{-1}$ for (a) $\Gamma_s = 0.1
 \Delta_0$ and (b) $\Gamma_s = 2 \Delta_0$. The dashed lines in the
 panel (a) correspond to the isolated superconductor $\Gamma_s = 0$.}
 \label{fig:kernel}
\end{figure*}

In the case $\Gamma_n = 0$
the self-energies of the
superconductor are determined by the equilibrium Green's functions of the
normal metal, and, assuming zero temperature limit we have:
\begin{equation}
    \begin{gathered}
            \check \Sigma^{s(R/A)}(\omega) = \pm i \Gamma_s\\ \check
            \Sigma^{s(K)}(\omega) = -
            2 i \Gamma_s \sign \omega \\
            \check{ G}_{0k}^{s(R/A)} = \left(\omega \pm i \Gamma_s - \check
        { H}_{0k}^s\right)^{-1}\\
        \check{ G}_{0k}^{s(K)} = \left[\check{
            G}_{0k}^{s(R)} -
        \check{ G}_{0k}^{s(A)}\right]\sign \omega
    \end{gathered}
\end{equation}
The self-energies are constant $\delta \Sigma^s = 0$ so the solution of the
equations~(\ref{eq:dyson-linear})
\begin{equation}
    \delta \breve{ G}_k^s(\omega, \omega') =
                    \breve{
                    G}_{0k}^s(\omega) \delta \breve H^s(\omega - \omega')
                    \breve{
                    G}_{0k}^s(\omega')
\end{equation}
using the selfconsistency equation~(\ref{eq:selfconsistency}) one can
obtain an expression for the kernel of the Higgs mode:
\begin{equation}
    K_{\Delta}(\omega) =   \frac{i \lambda \nu_s}{8 \pi }  \iint
    \Tr \left[\check \tau_1 \check{ G}_{0k}^{s(R)}(\omega + \omega')
        \check \tau_{1} {\check
    { G}}^{s(K)}_{0k}(\omega')
        + \check \tau_1 \check{ G}^{s(K)}_{0k}(\omega' + \omega) \check
    \tau_{1}
    \check{ G}^{s(A)}_{0k}(\omega')
\right]\;d\omega' \;d\xi_k^s \ .
\end{equation}
This integral diverges logarithmically, however it can be regularized using
the equilibrium selfconsistency equation for $\Delta_0$. Finally, one can
obtain the Eq.~(11) of the main text of the paper:
\begin{multline}
    \frac{1 - K_\Delta(\omega)}{\lambda \nu_s} =
       i \frac{\sqrt{4 \Delta_0^2 - (\omega + 2 i \Gamma_s)^2}}{2 \omega + 4 i \Gamma_s} \cdot \ln \left[
    \frac{2 \Delta_0^2 - (\omega + 2 i \Gamma_s) (\omega + i  \Gamma_s) +
        \sqrt{4 \Delta_0^2 - (\omega + 2 i  \Gamma_s)^2} \sqrt{\Delta_0^2 - ( \omega + i
        \Gamma_s)^2}}{2 \Delta_0^2 - i  \Gamma_s( \omega + 2 i  \Gamma_s) +
            \sqrt{4 \Delta_0^2 - (
        \omega + 2 i \Gamma_s)^2}\sqrt{\Gamma_s^2 + \Delta_0^2}}\right]+
        \\
   i \frac{ \sqrt{4 \Delta_0^2 - { \omega^2}}}{2  \omega}
        \ln \left[
            \frac{
                2 \Delta_0^2 + i  \Gamma_s  \omega + \sqrt{ \Gamma_s^2 +
                \Delta_0^2}
                \sqrt{4 \Delta_0^2 - \omega^2}
            }{
                2 \Delta_0^2-  \omega( \omega + i  \Gamma_s) + \sqrt{4
                    \Delta_0^2 -
                \omega^2}\sqrt{\Delta_0^2 - ( \omega + i  \Gamma_s)^2
                }
            } \cdot
            \frac{\omega + i \sqrt{4 \Delta_0^2 -  \omega^2}}
            { -\omega + i \sqrt{4 \Delta_0^2 -  \omega^2}}
        \right]
   \ .
\end{multline}
The plots of $[1 - K_\Delta(\omega)]^{-1}$ are shown in Fig.~\ref{fig:kernel}
for the cases of the (a) low  and (b) high tunneling rates.

\subsection{Derivation of the long time asymptotics of the order parameter}

\begin{figure}
 \centering
 \includegraphics[width=0.4\linewidth]{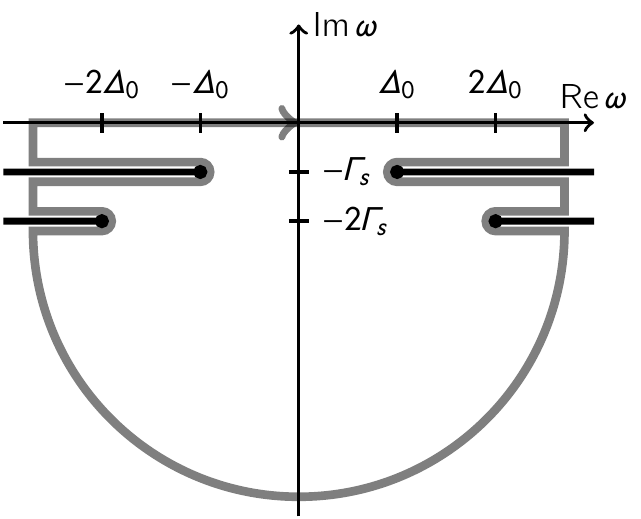}
 \caption{
 Area of analyticity of the kernel $K_\Delta$. The four black dots are
 the branch points $\omega = \pm \Delta_0 - i \Gamma_s$ and $\omega = \pm
 2 \Delta_0 - 2 i \Gamma_s$, the thick black lines are the branch cuts.
 The thick grey line shows the integration contour in the
 Eq.~(\ref{eq:contour}).
 }
 \label{fig:contour}
\end{figure}

We suppose that the external force $f(\omega)$ is a meromorphic function of the frequency
$\omega$ one can close up the integral over the real freqeuncies into a
contour integral as it is shown
in the Fig.~\ref{fig:contour}:
\begin{equation}
 \label{eq:contour}
 \frac{1}{2\pi} \oint\limits_{\mathcal C} \frac{f(\omega) e^{-i \omega
 t}\;d\omega}{1 - K_\Delta(\omega)} = -i \sum\limits_j
 \frac{e^{-i \omega_j t} \mathop{\mathrm{res}}_{\omega_j} f(\omega)}{1 -
 K_\Delta(\omega_j)} \
 ,
\end{equation}
where $\mathcal C$ is the integration contour, $\omega_j$ are the poles of the
external force
$f(\omega)$ within the contour $\mathcal C$ and $\mathop{\mathrm{res}}_{\omega_j} f(\omega)$ is the
residue of $f(\omega)$ at the pole $\omega_j$, thus, the integral in the
equation~(8) of the main paper
can be expressed as a sum of the integrals along the branch cuts and the terms
with the residues of the external force:
\begin{multline}
    \delta \Delta(t) = -i \sum\limits_j
    \frac{e^{-i \omega_j t} \mathop{\mathrm{res}}_{\omega_j} f(\omega)}{1 -
    K_\Delta(\omega_j)} -
    \frac{1}{2\pi} \left\{\int\limits_{\Delta_0 - i \Gamma_s}^{+\infty - i \Gamma_s}
    \left[
        \frac{f_1(\omega) e^{-i \omega t}}{1 - K_\Delta(\omega  - i 0)} -
        \frac{f_1(\omega) e^{-i \omega t}}{1 - K_\Delta(\omega + i 0)}
\right]\;d\omega + \right. \\
    \int\limits_{2 \Delta_0 - 2 i \Gamma_s}^{+\infty - 2 i \Gamma_s}
    \left[
        \frac{f_1(\omega) e^{-i \omega t}}{1 - K_\Delta(\omega  - i 0)} -
        \frac{f_1(\omega) e^{-i \omega t}}{1 - K_\Delta(\omega + i 0)}
        \right]\;d\omega +
    \int\limits_{-\infty - 2 i \Gamma_s}^{-2 \Delta_0 - 2 i \Gamma_s}
    \left[
        \frac{f_1(\omega) e^{-i \omega t}}{1 - K_\Delta(\omega  - i 0)} -
        \frac{f_1(\omega) e^{-i \omega t}}{1 - K_\Delta(\omega + i 0)}
    \right]\;d\omega + \\  \left.
    \int\limits_{-\infty - i \Gamma_s}^{-\Delta_0 - i \Gamma_s}
    \left[
        \frac{f_1(\omega) e^{-i \omega t}}{1 - K_\Delta(\omega  - i 0)} -
        \frac{f_1(\omega) e^{-i \omega t}}{1 - K_\Delta(\omega + i 0)}
\right]\;d\omega \right\}
\end{multline}
The integrals along the branch cuts can be evaluated approximately assuming
$f(\omega)$ are regular near the branch
points of $K_\Delta(\omega)$ and that the main contribution to these integrals comes
from the from the vicinity of the singularities.
The expansion of the kernel $K_\Delta(\omega)$ near its branch points at $\omega = \pm 2 \Delta_0 - 2 i \Gamma_s$
and $\omega = \pm \Delta_0 - i \Gamma_s$ in the limit $\Gamma_s \ll \Delta_0$
reads as follows:
\begin{gather}
         \omega = \Delta_0 - i \Gamma_s + \Omega: \frac{1 -
        K_\Delta}{\lambda \nu_s} \approx
        \frac{\pi}{2\sqrt{3}} + \frac{2 \Gamma_s
        \sqrt{-2 \Omega}}{\Delta_0^{3/2}} \ , \\
        \omega = -\Delta_0 - i \Gamma_s - \Omega:  \frac{1 -
        K_\Delta}{\lambda \nu_s} \approx
        \frac{\pi}{2\sqrt{3}} - \frac{2  \Gamma_s
        \sqrt{-2 \Omega}}{\Delta_0^{3/2}} \ , \\
        \omega = 2 \Delta_0 - 2 i  \Gamma_s +  \Omega:  \frac{1 -
        K_\Delta}{\lambda \nu_s} \approx
        \pi (1 + i) \sqrt{\frac{\Gamma_s}{\Delta_0}}+ \frac{\pi}{2}\sqrt{
        -\frac{\Omega}{\Delta_0}} \ , \\
         \omega = -2\Delta_0 - 2 i  \Gamma_s -  \Omega:  \frac{1 -
        K_\Delta}{\lambda \nu_s} \approx
        \pi (1 - i) \sqrt{\frac{\Gamma_s}{\Delta_0}} -
        \frac{\pi}{2}\sqrt{-\frac{ \Omega}{\Delta_0}}  \ .
\end{gather}
Using these expansions one can finally obtain an expression for the near
equilibrium oscillations of the superconducting gap:
\begin{multline}
    \delta \Delta (t) \approx -i \sum\limits_j
    \frac{e^{-i \omega_j t} \mathop{\mathrm{res}}_{\omega_j} f(\omega)}{1 -
    K_\Delta(\omega_j)} - \frac{1}{2\pi\lambda \nu_s} \left\{
        -\frac{12 \sqrt{2} \Gamma_s e^{-\Gamma_s t} \left[
                f_1(\Delta_0) e^{-i \Delta_0 t - i\frac{\pi}{ 4}} + f_1(-\Delta_0)
                e^{i \Delta_0 t + i \frac{\pi}{ 4}}
        \right]}{(\pi \Delta_0 t)^{3/2}} - \right. \\ -\left.
        \frac{2 \sqrt{\Delta_0} e^{-2 \Gamma_s t} \left[
                f_1(2\Delta_0) e^{-2 i \Delta_0 t + i \frac{\pi}{4}} + f_1(-2
                \Delta_0) e^{2i \Delta_0 t - i \frac{\pi}{4}}
    \right]}{(\pi t)^{1/2}}
\right\}
\end{multline}

\end{widetext}

\end{document}